\documentclass[twocolumn,showpacs,preprintnumbers,amsmath,amssymb]{revtex4}

\usepackage{graphicx}
\usepackage{dcolumn}
\usepackage{bm}

\begin{document}

\title{Solutions to the Jaynes-Cummings model without the rotating-wave
approximation}
\author{Qing-Hu Chen$^{1,2}$, Tao Liu$^{3}$, Yu-Yu Zhang$^{2,4}$, and Ke-Lin Wang$%
^{5}$}
\address{
$^{1}$ Center for Statistical and Theoretical Condensed Matter
Physics, Zhejiang Normal University, Jinhua 321004, P. R. China  \\
$^{2}$ Department of Physics, Zhejiang University, Hangzhou 310027,
P. R. China \\
$^{3}$Department of Physics, Southwest University of  Science and Technology, Mianyang 621010, P.  R.  China\\
$^{4}$ Center for Modern Physics, Chongqing University£¬ Congqing
400044, P. R.  China\\
$^{5}$Department of Modern Physics, University of  Science and
Technology of China,  Hefei 230026, P.  R.  China
 }
\date{\today}

\begin{abstract}
By using extended bosonic coherent states, the solution to the
Jaynes-Cummings   model without the rotating-wave approximation can
be mapped to that of a polynomial  equation with a single variable.
The solutions to this polynomial  equation can give all eigenvalues
and eigenfunctions of this model with all values of the coupling
strength and the detuning exactly, which can be readily applied to
recent circuit quantum electrodynamic systems operating in the
ultra-strong coupling regime.
\end{abstract}

 \maketitle

\address{
$^{1}$ Center for Statistical and Theoretical Condensed Matter
Physics, Zhejiang Normal University, Jinhua 321004, P. R. China  \\
$^{2}$ Department of Physics, Zhejiang University, Hangzhou 310027,
P. R. China \\
$^{3}$Department of Physics, Southwest University of  Science and Technology, Mianyang 621010, P.  R.  China\\
$^{4}$ Center for Modern Physics, Chongqing University£¬ Congqing
400044, P. R.  China\\
$^{5}$Department of Modern Physics, University of  Science and
Technology of China,  Hefei 230026, P.  R.  China
 }

The   Jaynes-Cummings (JC) model\cite{JC}  describes the interaction
of a two-level atom  (qubit) with a single bosonic mode. It is a
fundamental one in  quantum optics.  Based on the assumption of
nearresonance and relatively weak atom-cavity coupling,  the
rotating-wave approximation (RWA) is usually employed, and
analytically exact solution can be trivially obtained.

Recently, the JC model is closely related to condensed matter
physics. It can be realized in some solid-state systems recently,
such as one Josephson charge qubit coupling to an electromagnetic
resonator \cite{Wallraff}, the superconducting quantum interference
device coupled with a nanomechanical
resonator\cite{Chiorescu,squid}, and the most recently LC resonator
magnetically coupled to a superconducting
qubit\cite{Wang,Niemczyk,exp}. In traditional quantum optics where
the coupling between the two-level "natural" atom and the single
bosonic mode is quite weak, RWA is the most useful approximation.
However, in the circuit quantum electrodynamic (QED), the artificial
atoms may interact very strongly with on-chip resonant circuits\cite
{Niemczyk,exp,Deppe,Fink,Mooij}, the RWA can not describe well the
strong coupling regime\cite{Niemczyk}. Therefore, the JC model
without the RWA is the focus of current
interests\cite{Irish,chenqh,Werlang,liu,Hanggi,Nori,Hwang,chen10,chen10b,Hausinger,Casanova}.

However, due to the inclusion of the counter-rotating terms, the
Bosonic number is not conserved, the Bosonic Fock space has infinite
dimensions, so any solution without the RWA is highly nontrivial. In
the recent years, several non-RWA approaches have been proposed in
the Dicke model\cite{chenqh}, the  quantum Zeno effect\cite{zheng},
and the spin-boson model\cite{Yuyu}. Especially, by using extended
bosonic coherent states, the present authors  have solved the Dicke
model without the RWA exactly in the numerical sense\cite{chenqh}.
The most simple $N=1$ Dicke model is just the JC model. This
numerically exact solutions to the JC model are also described in
detail in Refs. \cite{chen10,chen10b}. To the best of our knowledge,
an analytical exact solutions are still lacking in the literature to
date.

In this paper, we propose a new method to solve exactly the JC model
without the RWA by means of extended bosonic coherent states.  The
correlations among bosons are added step by step until further
corrections will not change the results. Different from our previous
work where the pure coherent state are fixed, the eigenvalue
$\alpha$ of the pure coherent state in this paper is tunable. By
solving Schr$\stackrel{..}{o}$dinger equation, the expanded
coefficients can be expressed by $\alpha$ through a recurrence
relation. We then derived a polynomial equation with only single
variable $\alpha$. The solutions to this  polynomial equation can
give exactly all eigenfunctions and eigenvalues of the JC model
without the RWA with arbitrary parameters.

Without the RWA, the Hamiltonian of  a qubit  interacting with a
single bosonic mode reads  ($\hbar =1$)
\begin{equation}
H_0=\frac \Delta 2\sigma _z+\omega a^{\dagger }a+g\left( a^{\dagger
}+a\right) \sigma _x,
\end{equation}
where $a^{+}$ and $a$ are the bosonic annihilation and creation
operators of the cavity, $\Delta $ and $\omega $ are the frequencies
of the atom and cavity, $g$ is the atom-cavity coupling constant,
and $\sigma _k(k=x,y,z)$ is the Pauli matrix of the two-level atoms.
For convenience, we can write a transformed Hamiltonian with a
rotation around an $y$ axis by an angle $\frac \pi 2$
\begin{equation}
H=-\frac \Delta 2\sigma _x+a^{\dagger }a+g\left( a^{\dagger
}+a\right) \sigma _z  \label{Hamilton},
\end{equation}
in the unit of $\omega =1$.  Associated with this Hamiltonian is a
conserved parity $\Pi $, such that $ \left[ H,\Pi \right] =0$, which
is given by
\begin{equation}
\Pi =\sigma _x\exp \left( i\pi \widehat{N}\right),
\end{equation}
where $\widehat{N}=$ $a^{+}a$ is the bosonic number operator. $\Pi $
has two eigenvalues $\pm 1$, depending on whether the excitation
number is even or odd. So the system has the corresponding even or
odd parity.

We propose the following ansatz for the wavefunction
\begin{equation}
\left| \Psi \pm \right\rangle =\left( \
\begin{array}{l}
\sum_{n=0}^Mc_n\left( a^{+}\right) ^n\exp \left( \alpha a^{+}\right) \left|
0\right\rangle \\
\pm \sum_{n=0}^Mc_n\left( -a^{+}\right) ^n\exp \left( -\alpha a^{+}\right)
\left| 0\right\rangle
\end{array}
\right),  \label{wavefunction}
\end{equation}
where  $c_n$ is the expansion coefficient and $\alpha$ is the
eigenvalue of the coherent state and will be determined later. $\Psi
_{+}$ $\left( \Psi _{-}\right) $ is  the eigenfunction of the
even(odd) parity with the eigenvalue $+1(-1)$.   The
Schr$\stackrel{..}{o}$dinger equation then gives
\begin{eqnarray}
&&\left[ a^{+}a+g\left( a^{+}+a\right) \right] \sum_{n=0}^Mc_n\left(
a^{+}\right) ^n\exp \left( \alpha a^{+}\right) \left| 0\right\rangle
\nonumber \\
&&\mp \frac \Delta 2\sum_{n=0}^Mc_n\left( -a^{+}\right) ^n\exp \left(
-\alpha a^{+}\right) \left| 0\right\rangle  \nonumber \\
&=&E^{\pm }\sum_{n=0}^Mc_n\left( a^{+}\right) ^n\exp \left( \alpha
a^{+}\right) \left| 0\right\rangle.  \label{s_eq}
\end{eqnarray}
By using $[a,a^{+}]=1$, equating the coefficients of the terms of
$\left( a^{+}\right) ^m\exp \left( \alpha a^{+}\right) \left|
0\right\rangle $ on both sides, the above equations for both even
and odd parity can independently give the following identities for
any $m$
\begin{eqnarray}
&&\left( m+\alpha g\right) c_m+\left( \alpha +g\right) c_{m-1}+\left(
m+1\right) c_{m+1}  \nonumber \\
&&\mp \frac \Delta 2(-1)^m\sum_{j=0}^m\frac{(2\alpha
)^j}{j!}c_{m-j}=E^{\pm }c_m . \label{coe_eq}
\end{eqnarray}
By careful inspection of Eq. (\ref{wavefunction}), one can find that
$c_0$ is flexible in the Schr$\stackrel{..}{o}$dinger equation where
the normalization for the eigenfunction is not necessary, so we
select $c_0=1.0$. The linear term in $a^{+}$ in the Fock space can
be also determined by the value of $\alpha$ in the pure coherent
state $ \exp \left( \alpha a^{+}\right) \left| 0\right\rangle $. It
is useless that both $c_1$ and $\alpha$ contribute to the linear
term, so we can set $ c_1=0 $. Once the first two terms are fixed,
the coefficients of the other terms higher than $a^{+}$ should be
determined by solving Schr$\stackrel{..}{o}$dinger equation. The
constant term yields
\begin{equation}
E^{\pm }=\alpha g\mp \frac \Delta 2.  \label{zero}
\end{equation}
Inserting Eq. (\ref{zero}) into Eq. (\ref{coe_eq}), we have the following
recurrence equation
\begin{eqnarray}
c_{m+1} &=&-\frac 1{(m+1)g}[(m\pm \frac \Delta 2)c_m+\left( \alpha +g\right)
c_{m-1}  \nonumber \\
&&\mp (-1)^m\frac \Delta 2\sum_{j=0}^m\frac{(2\alpha
)^j}{j!}c_{m-j}]. \label{recurrence}
\end{eqnarray}
For $m=M$, the terms higher then $\left( a^{+}\right) ^M\exp \left(
\alpha a^{+}\right) \left| 0\right\rangle $ are neglected, we may
set $ c_{M+1}=0 $,  then we have
\begin{eqnarray}
&&\left( M\pm \frac \Delta 2\right) c_{_M}+\left( \alpha +g\right) c_{_{M-1}}
\nonumber \\
&&\mp (-1)^M\frac \Delta 2\sum_{j=0}^M\frac{(2\alpha
)^j}{j!}c_{_{M-j}} =0. \label{central}
\end{eqnarray}
Note that all coefficients $c's$ can be expressed by one variable
$\alpha$ through Eq. (\ref{recurrence}), so this is a one-variable
polynomial equation of degree  $M$.  This is the central equation in
this paper, which roots would give the exact solutions to the JC
model without the RWA.

To obtain the true exact results, in principle, the truncated number
$M$ should be taken to infinity. Fortunately, it is not necessary.
It is found that finite terms in state (\ref{wavefunction}) are
sufficient to give exact results in the whole coupling range.
Typically, the convergence is assumed to be achieved if the results
are determined within very small relative errors when the truncated
number $M$ increases further. We like to stress here that increasing
the value of $M$ would almost not bring additional effort to solve
Eq. (\ref{central}) in an ordinary PC. The precision for the results
is only limited to the machine accuracy for all cases tested.

To have a sense about this method, we perform the first-order
approximation (FOD) by considering $M=2$.  The only one coefficient
to be determined is $c_2$, which is easily obtained by the Eq. ({
\ref{recurrence}) $c_2=-\frac 1{2g}\left[ \left( 1\pm \Delta \right)
\alpha +g\right] $. The nonlinear equation then is given by
\[
\mp \Delta g\alpha ^2-\left( 1\pm \Delta \right) \alpha -g=0
\].
The value of $\alpha $ for the even parity $(+)$ and odd parity
$(-)$ are
\begin{eqnarray}
\alpha ^{+} &=&-\frac{\left( 1+\Delta \right) }{2\Delta g}\left(
1\pm \sqrt{
1-\frac{4\Delta g^2}{\left( 1+\Delta \right) ^2}}\right) , \\
\alpha ^{-} &=&\frac{\left( 1-\Delta \right) }{2\Delta g}\left( 1\pm
\sqrt{1+ \frac{4\Delta g^2}{\left( 1-\Delta \right) }}\right).
\end{eqnarray}
The corresponding energies are
\begin{equation}
E^{+}=-\frac{\left( 1+\Delta \right) ^2}{2\Delta }\left( 1\pm
\sqrt{1-\frac{ 4\Delta g^2}{\left( 1+\Delta \right) ^2}}\right)
+\frac 12, \label{energy_+}
\end{equation}
}
\begin{equation}
E^{-}=\frac{\left( 1-\Delta \right) ^2}{2\Delta }\left( 1\pm
\sqrt{1+\frac{ 4\Delta g^2}{\left( 1-\Delta \right) ^2}}\right)
+\frac 12.  \label{energy_-}
\end{equation}
Note that the lower eigenvalue for  the even parity $(+)$, i.e. Eq.
(\ref{energy_+} ) with plus sign in the RHS, increases with $g$. It
is physically unreasonable, because the ground-state(GS) energy
should decrease with the qubit-cavity coupling. The corresponding
coefficient $c_2$ is considerably larger than $c_0=1$, indicating
that the wavefunction is not converging. So this solution should be
omitted.

\begin{figure}[tbp]
\includegraphics[scale=0.6]{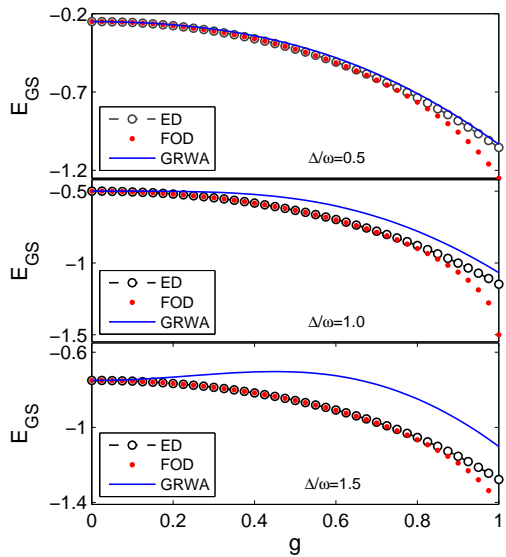}
\caption{(Color online) Comparisons of the GS  energies   by the
present FOD (solid circles), GRWA (solid curves) , and ED (open
circles) as a function of $g$ at $\Delta/\omega=0.5$ (top),
$\Delta/\omega=1$ (middle), and $\Delta/\omega=1.5$ (bottom). }
\label{comp3}
\end{figure}

The GS energies in the FOD with different ratios of $\Delta/\omega$
for $M=2$, i.e. Eq. (\ref{energy_+} ) with minus sign in the RHS,
are plotted in Fig. \ref{comp3} with solid circles. A generalized
RWA (GRWA) for JC model was performed  recently \cite{Irish} and the
derived expression for the energy levels  is significantly  more
accurate than that with RWA.  The GS energies by the GRWA with the
same ratios of $\Delta/\omega$ are also list Fig. \ref{comp3} with
solid curves for  comparison.  To show the accuracy, we also collect
the GS energies by the numerically exact diagonalization (ED) in
Bosonic Fock states $\frac{\left( a^{+}\right) ^m}{\sqrt{m!}}\left|
0\right\rangle $ with open circles.

It is interesting that  the GS energies by the present FOD are  much
more close to the numerical ED ones for the zero, positive and
negative large detuning. The relative difference for the GS energy
is less than $10^{-3}$ for $g\le 0.5$. One may deeply impressed by
this good agreement, because the present FOD is only a preliminary
approximation.  Note that the maximum coupling constant in the
ultrastrong-coupling regime in the  circuit QED   reported recently
\cite{Niemczyk,exp} is  around $g=0.12$. When only the GS energy is
concerned in these systems, the present very simple analytical
expression Eq.  (\ref{energy_+} ) with minus sign in the RHS should
be very helpful.

If we increase the truncated number $M$, we would obtain more
accurate energy levels.  On the basis of the Abel-Ruffini theorem
that the  general solution in radicals is impossible to polynomial
equations of degree five or higher\cite{Fraleigh}, the  most
accurate analytically expressions of the energy levels of the system
might be obtained by set $M=4$,  which is however not shown here due
to the complexity. Naively speaking, one could have 7 eigenstates
for $M=4$. Actually it is not that case. For the energy expression
Eq. (\ref{zero}), only real roots for $\alpha $ in Eq.
(\ref{central}) is reasonable, some complex roots should be omitted.

\begin{figure}[tbp]
\includegraphics[scale=0.6]{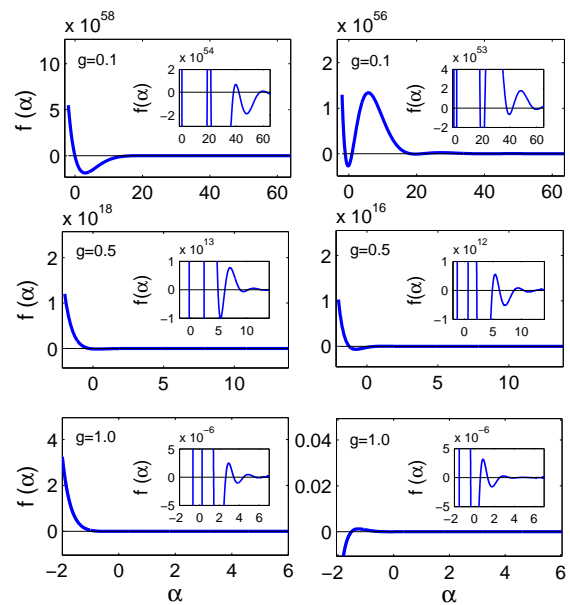}
\caption{(Color online) Schematic view of the solutions for the
one-variable polynomial  equation Eq. (\ref{central}) for $ g=0.1$
(top panel), $0.5$ (middle panel), and $1.0$ (bottom panel). The
left panel is for the even parity and the  right for the  odd
parity. The truncation numbers are $M=59$ for the even parity and
$M=60$ for odd parity. The insets show  the enlarged view. }
\label{solution}
\end{figure}

It is very  crucial to obtain the real roots of Eq. (\ref{central})
for sufficient large $M$ where the general solutions do not exist.
To achieve this goal, we plot a two dimensional diagram
$y=f(\alpha)$, where $f(\alpha)$ is just the LHS of  Eq.
(\ref{central}).  The schematic view of the solutions for the
one-variable polynomial equation Eq. (\ref{central}) for $g=0.1$,
$0.5$, and $1$ are presented in Fig. \ref{solution}.  The real roots
are just the crossing points of the curve $y=f(\alpha)$ and the
straight line $y=0$.

It is observed universally that if $M$ is a even (odd) number for
the even (odd) parity, the lowest eigenvalue for the even (odd)
parity is physically unreasonable, because the expansion
coefficients in the wavefunction (\ref{wavefunction}) are not
converging. This characteristic has shown up in the case of $M=2$
discussed above. Except this lowest eigenvalue, it will be confirmed
later that all other obtained eigenvalues of this model are true
eigenvalues of model. So for a given model parameters $g$ and
$\Delta$, we choose $M$ to be a odd (even) number for the even (odd)
parity so that the physically unreasonable solution does not appear.

In Fig. \ref{solution}, we choose  $M=59$ for the even parity and
$60$ for the odd parity, which is sufficient large to ensure the
convergence.  The value of $f(\alpha)$ is very large  in the small
$\alpha$ regime, and decreases quickly as $\alpha$ increases. The
real roots are clearly exhibited in this two-dimensional plot and
can be obtained  easily  by Maple in the practical calculation. The
number of the real  roots are considerably less than $M$. E.g. for
$g=0.1$, we obtain totally $27$ real roots with $13$ roots for the
even parity and $14$ roots for the odd parity,  which are
corresponding to $27$ energy levels.  Note that the roots for
$\alpha$ for the same energy levels generally decrease with the
coupling strength $g$. According to the energy levels Eq.
(\ref{zero}), this is physically reasonable, because the total
energy of this system should decrease with the coupling strength
$g$.

Once the roots are at hand, one natural question is whether the
energy levels and the wavefunctions corresponding to these roots are
really the exact ones to the JC model. Actually, for every model
parameters, we  increase $M$ by $2$  in each  step until the values
of the roots for $\alpha$ are not modified. Our criterion is that,
if the relative  difference for $\alpha$  is less than $10^{-8}$, we
think   $M$ is large enough to give the exact solutions to the JC
model.

\begin{figure}[tbp]
\includegraphics[scale=0.6]{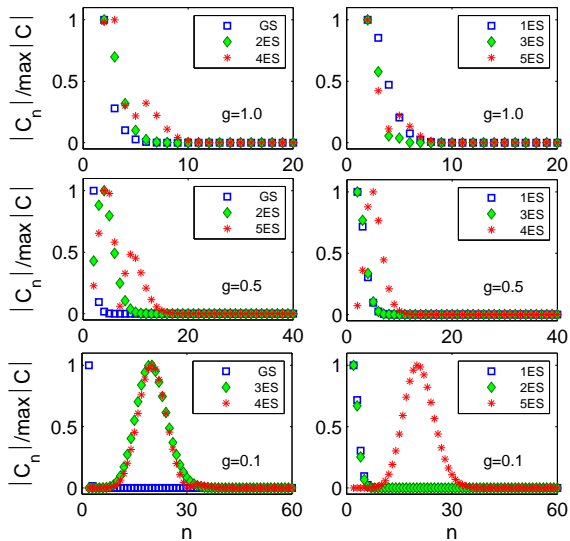}
\caption{(Color online) The absolute value of the coefficients $c_n$
for $g=0.1$ (bottom panel), $g=0.5$ (middle panel),  and $g=1$ (top
panel). The left panel is for the  even parity and right for odd
parity. The truncation numbers are $M=59$ for the even parity and
$60$ for the odd parity.} \label{coeff}
\end{figure}

The exact solutions are  also confirmed by the convergence of the
expanded coefficients in  the wavefunction (\ref{wavefunction}),
which are displayed in Fig. \ref{coeff}, where the absolute value of
coefficients $c_n$ normalized to the maximum value of $\{c\}$ for
$g=0.1$, $0.5$, and $1.0$ for the GS state and the first $5$ excited
states are plotted. It is very interesting that the coefficients
$c_n$ vanish after $n=40$ for $g=0.1$, $20$ for $g=0.5$, and $10$
for $g=1$. It follows that $M=40$ for $g=0.1$, $M=20$ for $g=0.5$,
and $M=10$ for $g=1$ are large enough to get the exact solutions for
these $6$ eigen states. Further increase of $M$ could not change the
wavefunction at all. In the practical calculations, if we choose the
truncation numbers $M=59$ for even parity and $60$ for odd parity,
we can get exact solutions for the first more than $20$ eigen states
for $ g \ge 0.1$, which might be practically very useful when
applied to the circuit QED system \cite{Niemczyk,exp}.

It is also interesting to note from   Fig. \ref{coeff} that the
necessary truncation number $M$ decrease as the coupling strength
increases, indicating that the present approach can be applied to
the JC model in the arbitrary ultra-strong coupling regime.  In
sharp contrast with the  present approach, in the numerically ED in
the bosonic Fock states, the dimension the truncated  subspace
increases considerably with the coupling constant $g$, due to more
photons are involved.  In the present approach, a infinite photons
have been already included in the bosonic coherent states in the
wavefunction (\ref{wavefunction}).

In the strong coupling limit, the first term in the JC Hamiltonian
(\ref{Hamilton}) can be neglected, so we have
\begin{equation}
H_0=a^{+}a+g(a+a^{+})\sigma _z.  \label{scl}
\end{equation}
By introducing the new operators\cite{chen10}
\begin{equation}
A=a+g,B=a-g  \label{newoperators}
\end{equation}
we can observe that the linear term for the bosonic operator is
removed, and only the number operators $A^{+}A$ and $B^{+}B$ are
left. Therefore we can readily obtain the eigenfunctions
\begin{equation}
\left| m\right\rangle =\left(
\begin{array}{l}
\left| m\right\rangle _A \\
\pm (-1)^m\left| m\right\rangle _B\label{wf_scl}
\end{array}
\right).
\end{equation}
with
\begin{equation}
\left| m\right\rangle _A=\frac{\left( a^++g/\omega \right) ^m}{\sqrt{m!}}%
\left| 0\right\rangle _A,\left| m\right\rangle _B=\frac{\left(
a^+-g/\omega \right) ^m}{\sqrt{m!}}\left| 0\right\rangle _B,
\end{equation}
and the eigenvalues $E_m^{\pm}=m-g^2$ for the $m$ state. Note that
the eigenstates are twofold degenerate in the strong coupling limit.
Comparing with the present wavefunction (\ref{wavefunction}), we can
find that the truncation number is just fixed to be $m$ for the
$m$th  excited state. In the strong coupling regime, as the coupling
strength $g$ increases, only a fewer terms are  needed to add to the
$m$th wavefunction Eq. (\ref{wf_scl}) for the $m$th state. The
stronger the coupling strength is, the fewer additional terms are
needed, which highlights the  advantage of this method.

Finally, to show the exact nature of the solutions directly, we
compare the present analytically exact solutions with those by
numerical ED in Bosonic Fock states in Tables I, II and III. We only
list the energies of the GS and the first 8 excited states for three
typical detunings $\Delta/\omega=1, 0.5$, and $1.5$ at different
coupling strength where the RWA is usually invalid. It is very
important to observe that the present results are the same as those
by ED in all cases. We should pointed out here that in the numerical
ED, the dimension of the truncated subspace is up to $10^5$ at
strong coupling, more computer memory is needed to store the matrix
elements, and several hours of CPU time is taken to perform the
diagonalization. In the present approach, the dimension of the
truncated subspace is only around $60$, and the results are obtained
within one second in any PC. More importantly, we only need to solve
one polynomial equation with a single-variable. To the best of our
knowledge, this  is  the first analytically exact solution to the JC
model without the RWA for arbitrary  coupling strengths and
detunings, which is very simple and can be easily employed.

\section*{ACKNOWLEDGEMENTS}

This work was supported by National Natural Science Foundation of
China, National Basic Research Program of China (Grant Nos.
2011CBA00103 and 2009CB929104), Zhejiang Provincial Natural Science
Foundation under Grant No. Z7080203, and Program for Innovative
Research Team in Zhejiang Normal University.

\begin{table}[h]
\caption{The first 9 low-lying energy levels $E_{i},
(i=0,1,2,...,8)$ in the case of  resonance $\Delta/\omega=1$.}
\begin{tabular}{p{1cm} p{2.5cm}  p{2.5cm} p{2.5cm} p{2.5cm}  p{2.5cm} p{2.5cm} }
 \hline
 \hline
  $E_{i}$  & present & ED & present & ED & present & ED     \\
 & g=0.1 & g=0.1 & g=0.5 & g=0.5 & g=1.0 & g=1.0     \\
 \hline

  $E_{0}$  & -0.505012531 & -0.505012531 & -0.633294235 & -0.633294235 & -1.14794573 & -1.14794573   \\
  $E_{1} $  &0.395102298 & 0.395102298 & -0.120023834 & -0.120023834 & -1.01017830 & -1.01017830     \\
  $E_{2}$  &0.594847069 & 0.59484707 & 0.695393717 & 0.695393717 & -0.231722500 & -0.231722500     \\
  $E_{3}$  &1.35388915 & 1.35388915 & 0.82530520 & 0.82530520   & 0.133435454 & 0.133435454      \\
  $E_{4}$  &1.63600849 & 1.63600849 & 1.58705308 & 1.58705309 & 0.927043866 & 0.927043866     \\
  $E_{5}$  &2.32238587 & 2.32238587 & 1.93553948 & 1.93553948 & 1.10480946 & 1.10480946     \\
  $E_{6}$  &2.66745885 & 2.66745885 & 2.54858735 & 2.54858735 & 1.84278099 & 1.84278099     \\
  $E_{7}$  &3.29593219 & 3.29593219 & 2.94783100 & 2.94783100 & 2.14361945 & 2.14361945     \\
  $E_{8} $  &3.69385838 & 3.69385838 & 3.90961541 & 3.90961541 & 2.94392772 & 2.94392772     \\
 \hline
 \hline
\end{tabular}
\end{table}

\begin{table}[h]
\caption{The first 9 low-lying energy levels $E_{i},
(i=0,1,2,...,8)$
 for large detuning $\Delta/\omega=0.5$.}
\begin{tabular}{p{1cm} p{2.5cm}  p{2.5cm} p{2.5cm} p{2.5cm}  p{2.5cm} p{2.5cm} }
 \hline
 \hline
  $E_{i}$  & present & ED & present & ED & present & ED     \\
 & g=0.1 & g=0.1 & g=0.5 & g=0.5 & g=1.0 & g=1.0     \\
 \hline
  $E_{0}$  & -0.256681491 & -0.256681491 & -0.425996230 & -0.425996229 & -1.05412447 & -1.05412447     \\
  $E_{1}$  & 0.756227984 & 0.756227984 & -0.135825244 & -0.135825244 & -0.986090220 & -0.986090224     \\
  $E_{2}$  & 1.21812615 & 1.21812615 & 0.741771730 & 0.741771731 & -0.110627502 & -0.110627502     \\
  $E_{3}$  & 1.76842053 & 1.76842053 & 0 .760071830 & 0.760071831 & 0.0868714030 & 0.0868714028     \\
  $E_{4}$  & 2.20649908 & 2.20649908 & 1.67427980 & 1.67427980 & 0.967028390 & 0.967028388     \\
  $E_{5}$  & 2.77998652 & 2.77998652 & 1.83251690 & 1.83251690 & 1.04043102 & 1.04043102     \\
  $E_{6}$  & 3.19542589 & 3.19542589 & 2.64925491 & 2.64925491 & 1.92078003 & 1.92078003     \\
  $E_{7}$   & 3.79099791 & 3.79099791 & 2.85156863 & 2.85156863 & 2.07700324 & 2.07700324     \\
  $E_{8}$  & 4.18484822 & 4.18484822 & 3.65576197 & 3.65576197 & 2.96634670 & 2.96634670     \\
 \hline
 \hline
\end{tabular}
\end{table}

\begin{table}[h]
\caption{The first 9 low-lying energy levels $E_{i},
(i=0,1,2,...,8)$ for  large detuning $\Delta/\omega=1.5$.}
\begin{tabular}{p{1cm} p{2.5cm}  p{2.5cm} p{2.5cm} p{2.5cm}  p{2.5cm} p{2.5cm} }
 \hline
 \hline
  $E_{i}$  & present & ED & present & ED & present & ED     \\
 & g=0.1 & g=0.1 & g=0.5 & g=0.5 & g=1.0 & g=1.0     \\
 \hline
   $E_{0}$   & -0.754009629 & -0.754009629 & -0.856475589 & -0.856475589 & -1.27755156  & -1.27755156     \\
    $E_{1}$   & 0.223060923 & 0.223060923 & -0.203235835 & -0.203235837 & -1.06857920 & -1.06857920     \\
   $E_{2}$  & 0.768900004 & 0.768900004 & 0.600664430  & 0.600664428 & -0.357181745 & -0.357181745     \\
    $E_{3}$  & 1.20197785 & 1.20197785 & 0.959669469 & 0.959669469 & 0.130321638 & 0.130321638     \\
   $E_{4}$   & 1.78994281 & 1.78994281 & 1.47931569 & 1.47931569 & 0.871878630 & 0.871878627     \\
    $E_{5}$   & 2.18237074 & 2.18237074 & 2.04745904 & 2.04745904 & 1.19903842 & 1.19903842     \\
    $E_{6}$  & 2.80950833 & 2.80950833 & 2.45896800 & 2.45896801 & 1.76995826 & 1.76995826     \\
    $E_{7}$   & 3.16398011 & 3.16398011 & 3.00884499 & 3.00884499 & 2.18522105 & 2.18522105     \\
    $E_{8}$  & 3.82785600 & 3.82785600 & 3.53494142 & 3.53494142 & 2.95127586 & 2.95127586     \\
 \hline
 \hline
\end{tabular}
\end{table}
\end{document}